\documentclass{PoS}
\usepackage{float,amsmath}

\title{Simulation study for the proposed wide field-of-view gamma-ray detector array ALTO}

\ShortTitle{Simulation study of ALTO}

\author{\speaker{Satyendra Thoudam}$^{,1}$, Yvonne Becherini$^1$, Michael Punch$^{1,2}$\\
        
$^{1}$ Department of Physics and Electrical Engineering, Linnaeus University, 35195 Växjö, Sweden\\
$^{2}$ Astroparticule et Cosmologie (APC), CNRS, Universite Paris 7 Denis Diderot, 10,
	rue Alice Domon et Leonie Duquet, F-75205 Paris Cedex 13, France\\
        E-mail: \email{satyendra.thoudam@lnu.se}}

\abstract{ALTO is a wide field-of-view air shower detector array for very-high-energy (VHE) gamma-ray astronomy, proposed to be installed in the Southern Hemisphere at an altitude of $\sim\,5.1$~km above sea level. The array will use water Cherenkov detectors, as in the HAWC observatory, but combined with scintillator detectors, to detect air showers induced by VHE gamma rays in the atmosphere. It is being designed to attain a lower energy threshold, better energy and angular resolution, and improved sensitivity. The array will consist of $\sim\,1250$ small-sized ($3.6$~m diameter) detector units distributed over a circular area of $\sim\,160$~m in diameter. Each detector unit will consist of a water Cherenkov detector and a liquid scintillation detector underneath which will preferentially identify muons, facilitating the background (cosmic ray) rejection, thereby improving the sensitivity. The background rejection will be further enhanced by the close-packed arrangement and the small size of the detectors which will allow a fine sampling of air-shower footprints at the ground. In this contribution, we present the Monte-Carlo simulation of the experiment performed using CORSIKA and GEANT4 simulation packages. The expected performance of the array in terms of reconstruction accuracies of the shower core and arrival direction, as well as preliminary estimate of the trigger energy threshold after preliminary selection cuts for a point-like gamma-ray source are presented.}

\FullConference{35th International Cosmic Ray Conference --- ICRC2017\\
		10--20 July, 2017\\
		Bexco, Busan, Korea}

\begin{document}

\section{Introduction}
In the last decade, Imaging Atmospheric Cherenkov Telescopes (IACTs) such as H.E.S.S., MAGIC and VERITAS have made significant progress in very-high-energy (VHE) gamma-ray astronomy with the discovery of close to $200$ VHE gamma-ray sources \cite{Holder2015}. Although IACTs are highly sensitive instruments, there are limitations due to their narrow field-of-view and limited observing time, as they can be operated mostly during dark moonless nights. Mainly due to these limitations, transients and extended sources largely remain an unchartered territory in VHE gamma rays. These limitations are overcome with particle detector arrays such as Water Cherenkov Detector (WCD) arrays, but at some expense of sensitivity as well as angular and energy resolutions. WCD arrays detect VHE gamma rays via the detection of Cherenkov light produced by air shower particles as they traverse through the water volume. HAWC experiment in Mexico, with $\sim\,300$ water tanks, is currently the most sensitive experiment based on the WCD technique dedicated for VHE gamma-ray astronomy \cite{HAWC}.

ALTO is a VHE gamma-ray experiment based on a WCD array, but with the addition of a scintillator detector layer, which is proposed to be installed in the Southern Hemisphere at an altitude of $\sim\,5.1$~km above sea level. It is being designed to attain a lower energy threshold, better angular resolution and improved sensitivity with respect to HAWC. In the following, we present a brief description of the ALTO array, and describe the Monte-Carlo simulation that has been carried out to study the expected performance of the array such as the reconstruction accuracies of air shower parameters and the energy threshold for a point-like gamma-ray source.

\section{ALTO array and detector design}
\begin{figure}
\centering
\includegraphics[width=0.45\textwidth,height=0.45\textwidth]{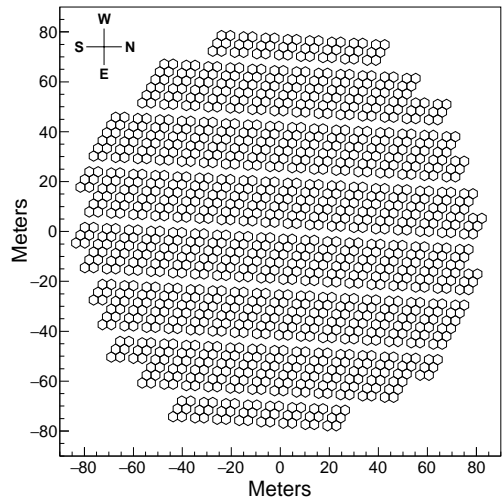}
\caption{\label{alto-array} \small{Layout of the ALTO array.}}
\end{figure}

\begin{figure}
\centering
\includegraphics[width=0.54\textwidth,height=0.397\textwidth]{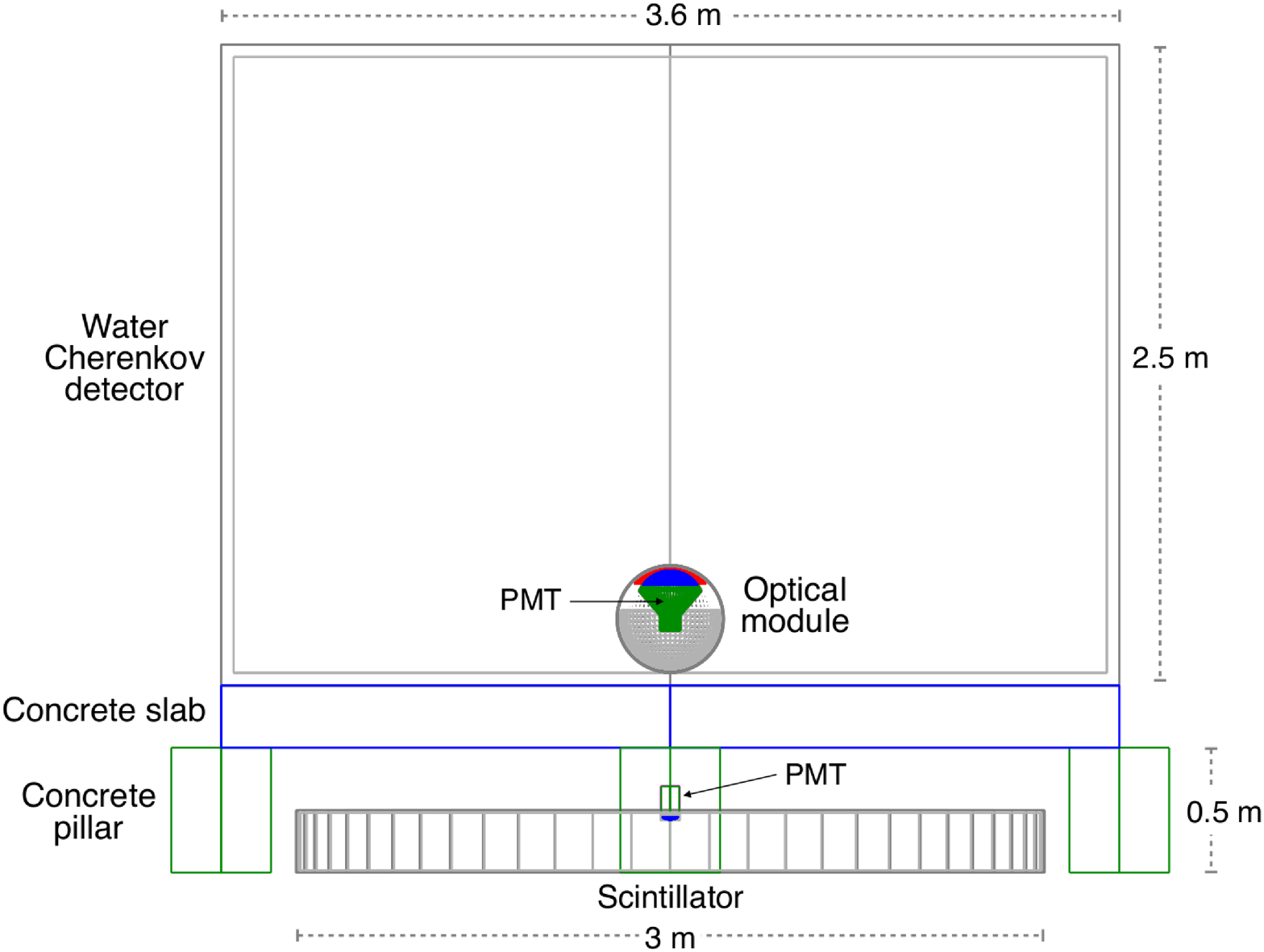}
\includegraphics[width=0.35\textwidth,height=0.392\textwidth]{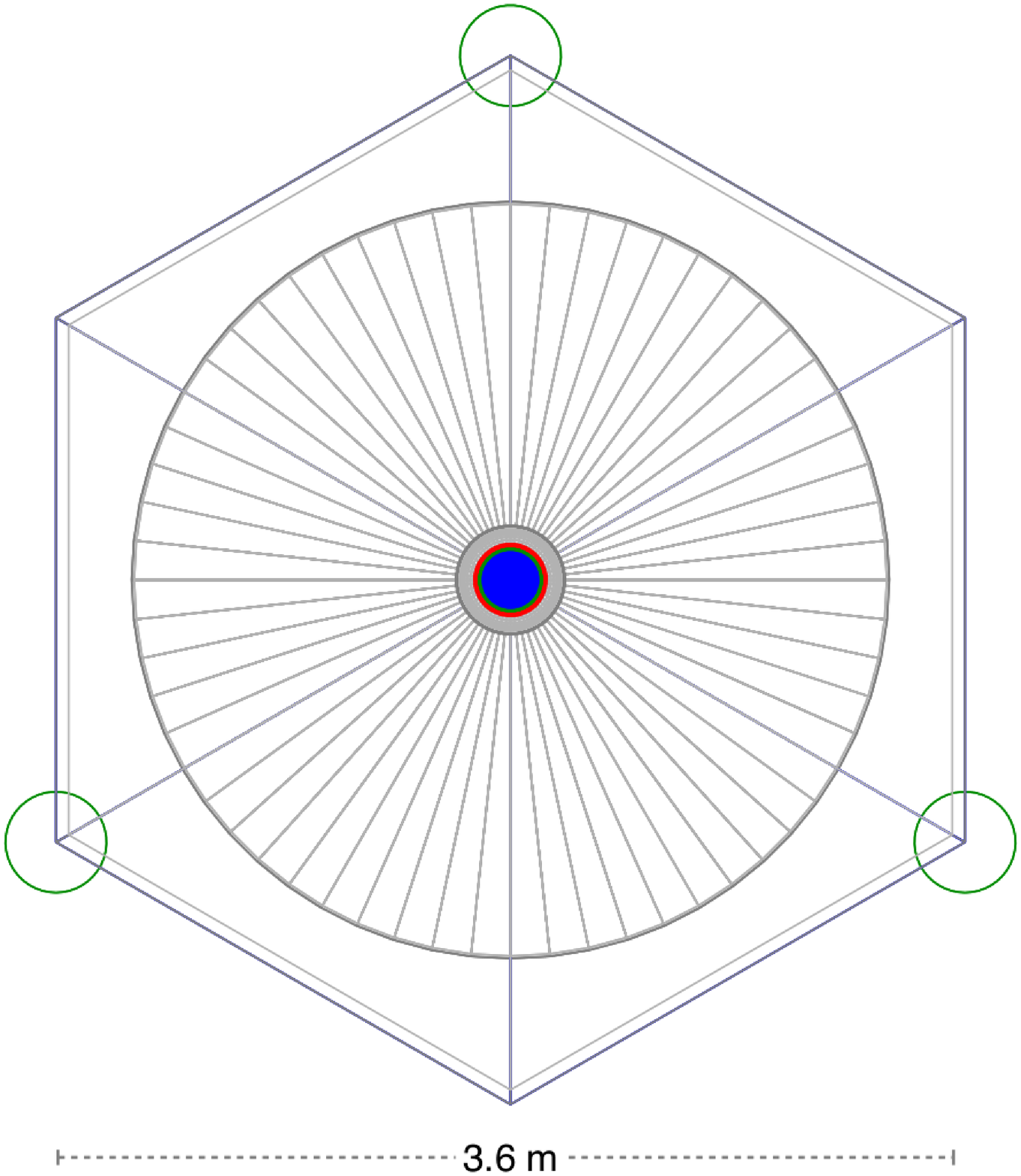}
\caption{\label{alto-unit} \small{Schematic of an ALTO detector unit. \textit{Left panel:} Side view. \textit{Right panel:} Arial view. The blue patch inside the optical module represents photocathode, and the the red colour represents optical glue that sticks the photomultiplier tube (green) into the optical module.}}
\end{figure}

The current design of ALTO consists of 1242 detector units distributed in a close arrangement within an area of 160~m across (see Figure \ref{alto-array}). Each detector unit consists of two segments: a hexagonal WCD and a cylindrical liquid scintillator detector (SD) underneath, as depicted in Figure \ref{alto-unit} along with their specifications. The dimension of WCD is almost half that of the HAWC experiment. This, together with the closed packed arrangement, will allow a finer sampling of air shower footprints, improving the discrimination between gamma-ray and cosmic-ray initiated showers, and thereby the sensitivity.

The sensitivity will be further improved using signals from the SDs, which will serve as muon detectors. As muons are present primarily in air showers initiated by cosmic rays, their detection is a powerful tool to identify the type of the incident primary particle. Electron, positrons and low-energy gammas in the air showers will mostly get attenuated in the WCD as they traverse the water volume. Muons will mostly be able to penetrate through, and reach the SD. The hexagonal shape of the WCD allows a more compact arrangement between adjacent tanks, providing a better shielding of the SDs from electrons, positrons and low-energy gammas. The WCDs will rest on concrete table which will further improve shielding of the SDs.

The WCD tank will be made of carbon fibre or similar composite material for strength and lightness. Cherenkov light produced by cascade particles in the water tank will be detected with a $10$-inch Hamamatsu photomultiplier tube (PMT) R7081-20 mounted inside an optical module placed at the bottom of the tank. As of now, the plan is to reuse optical modules from the ANTARES neutrino telescope \cite{Antares}, as  depicted also in Figure \ref{alto-unit} (left). The inner walls of the ALTO water tank will be made as much non-reflective as possible so as to preserve the timing information of the arriving particles. The SD tank will be made of aluminium, and the scintillating material will be an organic liquid Linear Alkyl Benzene mixed with small quantities of wavelength shifter powders PPO and POPOP as used for the ANTARES surface array\footnote{J. P. Ernenwein, private communication.}. Scintillation light, mostly in the visible range, is produced when a charged particle passes through the liquid scintillator. The light will be collected with a smaller $3$-inch Hamamatsu PMT R6091 placed at the top of the scintillator tank as depicted in Figure \ref{alto-unit} (left). Contrary to the water tank, the inside of the scintillator tank will be treated or will have a tyvek coating to make it as reflective as possible so as to allow maximum collection of scintillation light. It may be noted that only signals from the water tank will be used for timing studies. For more details on the technical design of ALTO, see Ref. \cite{Yvonne}.

\section{Air shower and detector simulations}
Air showers are simulated using the CORSIKA simulation package (version $7.4387$) \cite{Heck}. The interactions of hadronic particles in the Earth's atmosphere are treated using QGSJET-II-04 at high energies and FLUKA  for energies below $200$~GeV. The electromagnetic interactions are treated with EGS4. The observation level for ALTO is set to $5.1$~km above sea level, which is the altitude of the ALMA site in Chile \cite{ALMA}, and areas near the LLAMA site in Argentina \cite{LLAMA}.

The generated air shower particles are fed into a detector simulation code specifically developed for ALTO, based on the GEANT4 package \cite{GEANT4}. The code includes the design and geometry of each of the detector components along with all material properties such as density and refractive index as well as reflectivity, absorption and scattering coefficients of photons as function of wavelength. The code also include all important electromagnetic processes such as photoelectric effect, Compton scattering and pair production for gammas, multiple scattering, ionisation and bremsstrahlung for charged particles as well as decay processes for unstable particles. Optical processes such as emission, scattering and absorption of Cherenkov and scintillation photons are also included. For every air shower particle that passes through the water/scintillator tank, the code calculates the signal pulse (number of photoelectrons in bins of $0.5$~ns) generated at the photocathode of the PMT. An additional routine then translates the signal into a final waveform that takes into account the gain variation and transit time spread of the PMT.

\begin{figure}
\centering
\includegraphics[width=0.5\textwidth,height=0.42\textwidth]{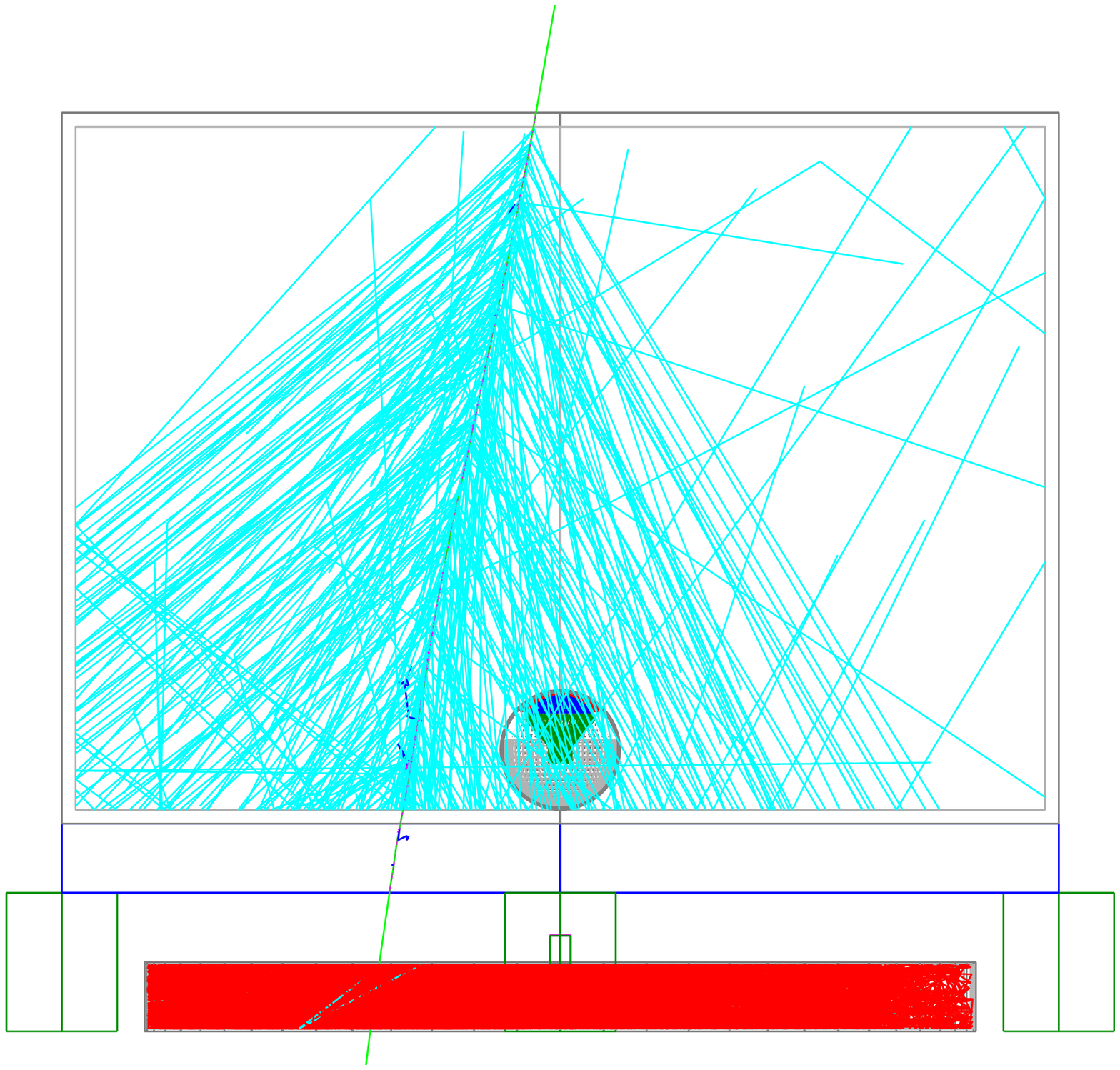}
\hspace{-0.37cm}
\includegraphics[width=0.5\textwidth,height=0.42\textwidth]{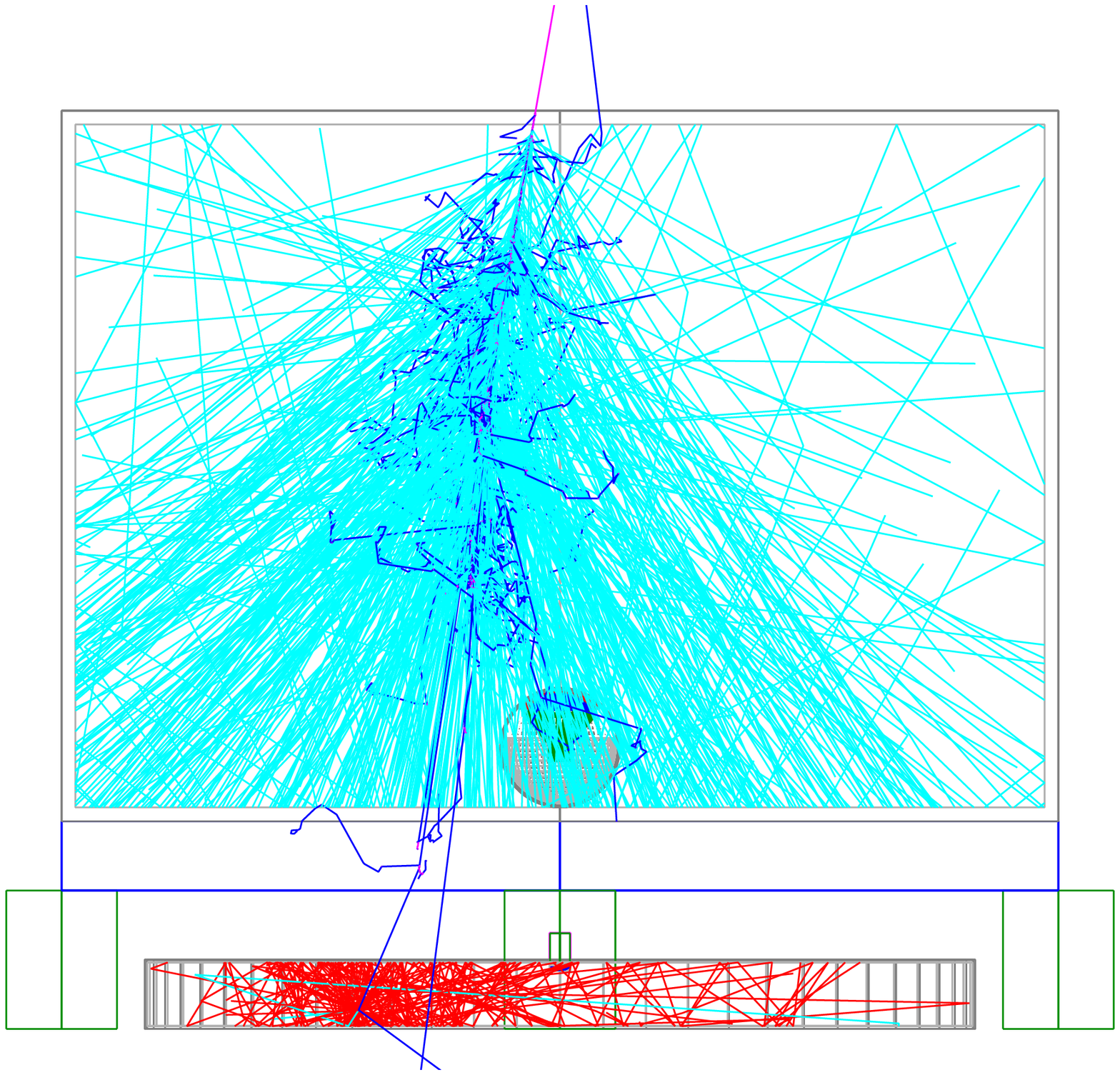}\\
\vspace{0.2cm}
\includegraphics[width=0.47\textwidth]{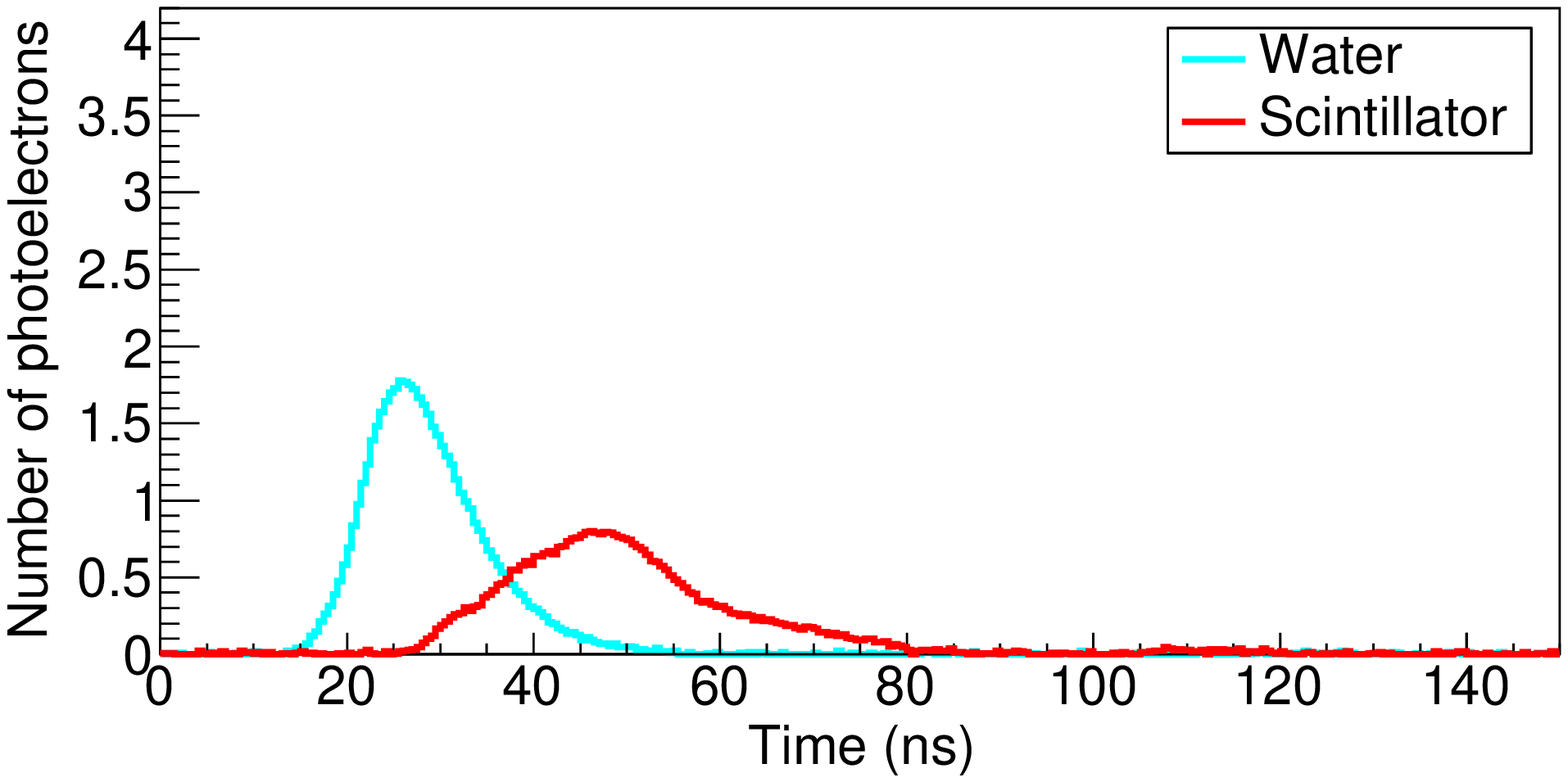}
\hspace{0.2cm}
\includegraphics[width=0.47\textwidth]{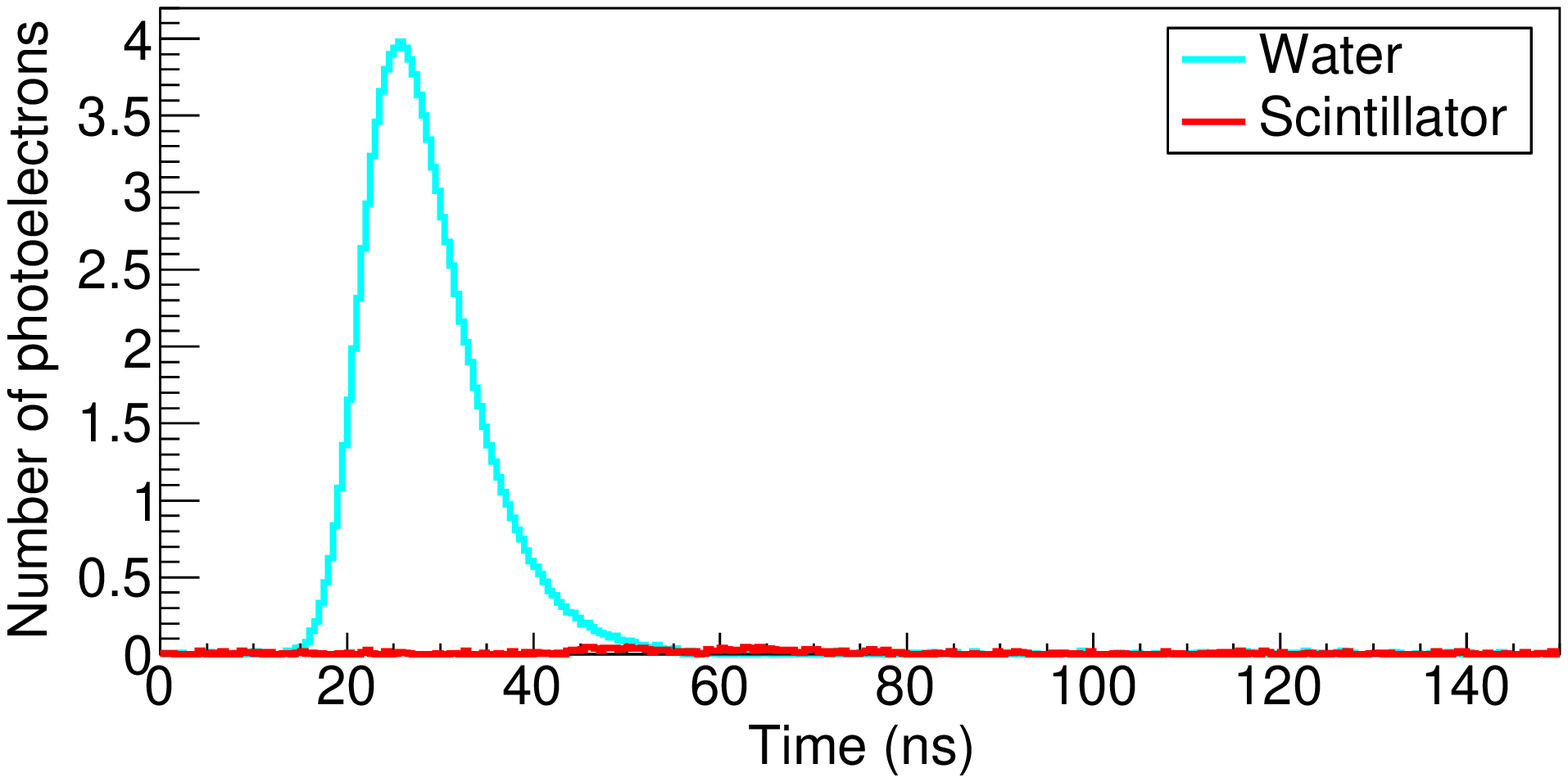}
\caption{\label{muon-electron-event} \small{\textit{Top panels:} A muon (\textit{top-left}) and an electron (\textit{top-right}) of 1~GeV energies hitting an ALTO detector unit. The muon (green track) passes straight through both the water and the scintillator tanks, while the electron (magenta track) develops a cascade in the water tank, and hardly manages to reach the scintillator tank. Different colours represent different particles/photons: muon (green), electron/positron (magenta), gammas (blue), Cherenkov photons (aqua), and scintillation photons (red). The blue track observed above the water tank in the case of electron is a gamma ray which escapes upward from the water tank. \textit{Bottom panels:} Signals recorded by the PMTs in the water (aqua) and the scintillator (red) tanks for the two cases.}}
\end{figure}

As Cherenkov and scintillation photons are produced in order of a few 100 thousands to millions in each detector volume, it is almost impossible to track all of them in the simulation. For optical photons in the WCDs, only those which would hit the optical module are allowed to be tracked by GEANT4, and the rest are killed at the point of emission. For photons produced inside the scintillator, they are not tracked in the simulation. Instead, we use a library of PMT signal waveforms for electrons, gammas and muons generated as a function of energy, position and incident angle on the scintillator.

Figure \ref{muon-electron-event} shows example of a muon and an electron of 1~GeV energies passing through a detector unit. The muon penetrates straight through both the water and the scintillator tanks. The electron, on the other hand, generates a cascade of secondary particles inside the water tank similar to an air shower, and almost get fully attenuated. A few very low-energy secondaries manage to cross the water tank, most of which gets absorbed in the concrete slab and ultimately only a tiny fraction manage to reach the scintillator producing a negligible signal. The signals (final waveforms) at the PMTs are shown in the corresponding bottom panels in Figure \ref{muon-electron-event}.

\section{Reconstruction of air shower parameters}
For each air shower, the signal arrival time\footnote{The signal arrival time is set at the time when the waveform passes 0.2 photoelectrons.} and the charge (integral of the signal waveform) recorded in each detector are determined. The relative signal arrival times between the WCDs are used to reconstruct the arrival direction of the shower, and the charge collected are used to determine the position of the shower axis (shower core). The reconstruction is performed mainly in two steps. In the first step, the arrival direction is reconstructed assuming a conical shape of the shower wavefront. In the second step, the shower core is reconstructed by fitting the two-dimensional distribution of the recorded charges (projected onto a plane perpendicular to the reconstructed direction, hereafter "shower plane") with the NKG (Nishimura-Kamata-Greisen) lateral distribution function given by \cite{Kamata, Greisen},
\begin{equation}
\label{eq:lateral-density}
\centering
\rho(r)=N_\mathrm{ch} C(s) \left(\frac{r}{r_\mathrm{M}}\right)^{s-2}\left(1+\frac{r}{r_\mathrm{M}}\right)^{s-4.5},
\end{equation}
where $\rho(r)$ represents the charge density as a function of distance $r$ from the shower core measured in the shower plane, $N_\mathrm{ch}$ is the total charge which corresponds to the number of particles contained in the air shower at the ground, $s$ is shower age, $r_\mathrm{M}$ is the radius parameter, and the function $C(s)$ is given by,
\begin{equation}
\label{eq-Cs}
\centering
C(s)=\frac{\Gamma(4.5-s)}{2\pi r_\mathrm{M}^2 \Gamma(s)\Gamma(4.5-2s)} .
\end{equation}
The reconstruction procedure described above is repeated iteratively with the output of each iteration taken as the starting values for the next iteration. For the analysis presented here, we use three iterations since this converges quickly.
\begin{figure}
\centering
\includegraphics[width=0.48\textwidth]{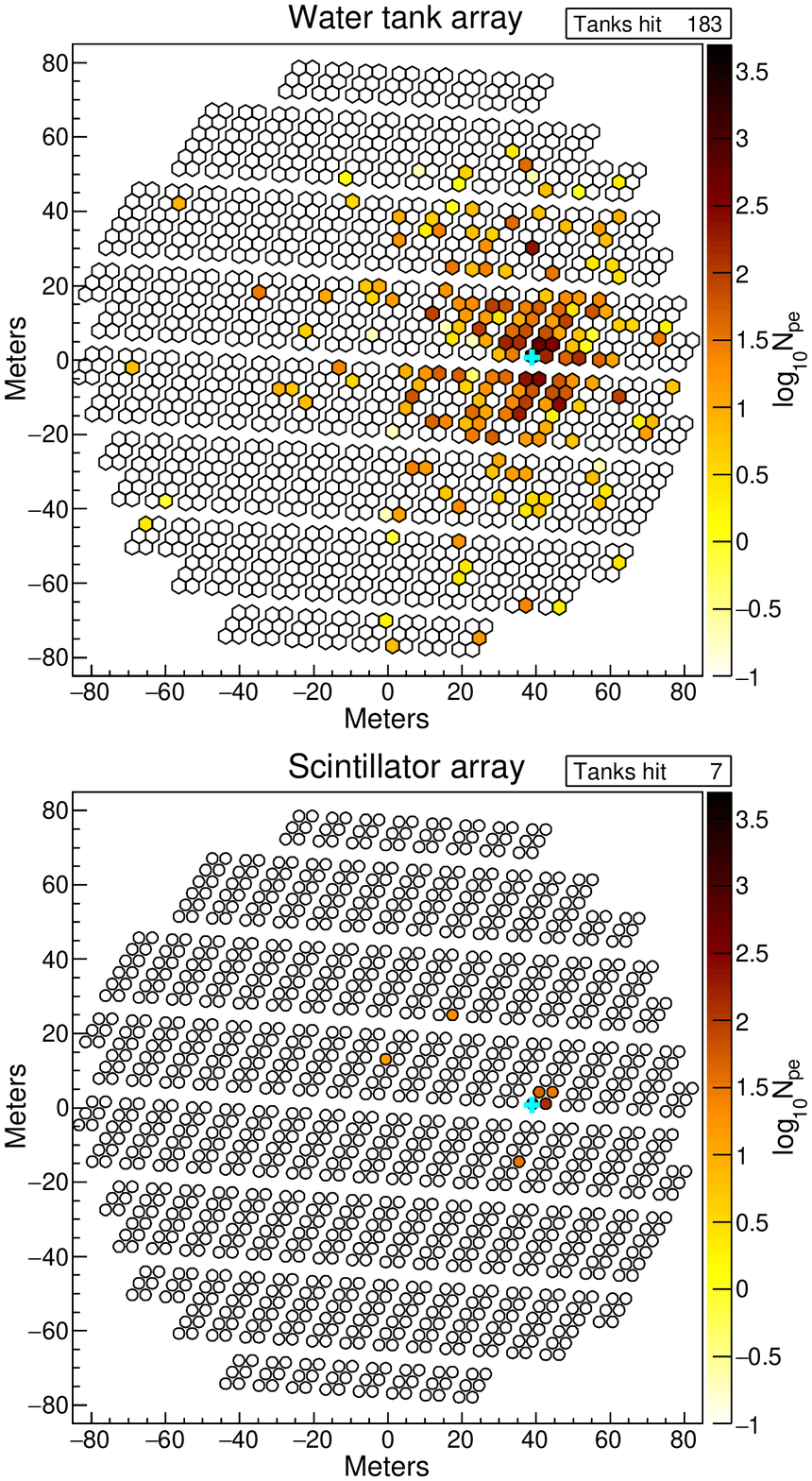}
\includegraphics[width=0.48\textwidth]{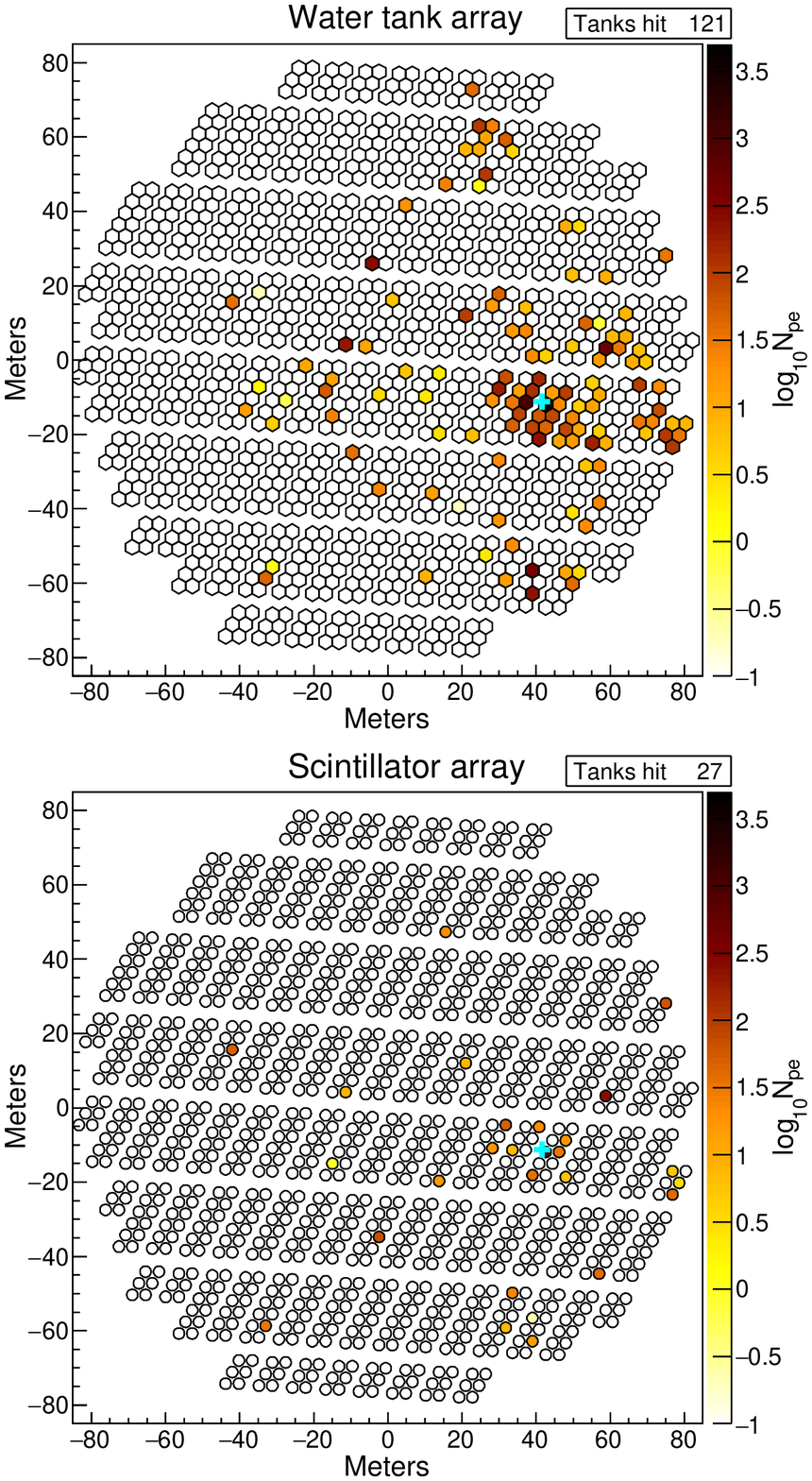}
\caption{\label{gamma-proton-event} \small{A simulated gamma-ray (\textit{left}) and a cosmic-ray proton (\textit{right}) event of energy $1$~TeV as observed with the ALTO array. \textit{Top panel:} Water tank array. \textit{Bottom panel:} Scintillator array. The colour scale represents the charge (total number of photoelectrons $\mathrm{N_{pe}}$) recorded in the tanks in logarithmic scale. The "+" sign (aqua colour) represents the position of the reconstructed shower core.}}
\end{figure}

Figure \ref{gamma-proton-event} shows examples of a simulated gamma-ray and proton events of 1~TeV energy as observed with ALTO. The reconstructed shower core is represented by the "+" sign. For the signals at the WCDs, the gamma-ray event shows a more compact and smoother distribution, while the proton event exhibits several spots of strong signals away from the reconstructed core which are mainly due to the presence of muons. The difference is seen to be even more pronounced at the SD level. 

\section{Preliminary results on the array performance}
\begin{figure}
\centering
\includegraphics[width=0.48\textwidth,height=0.29\textwidth]{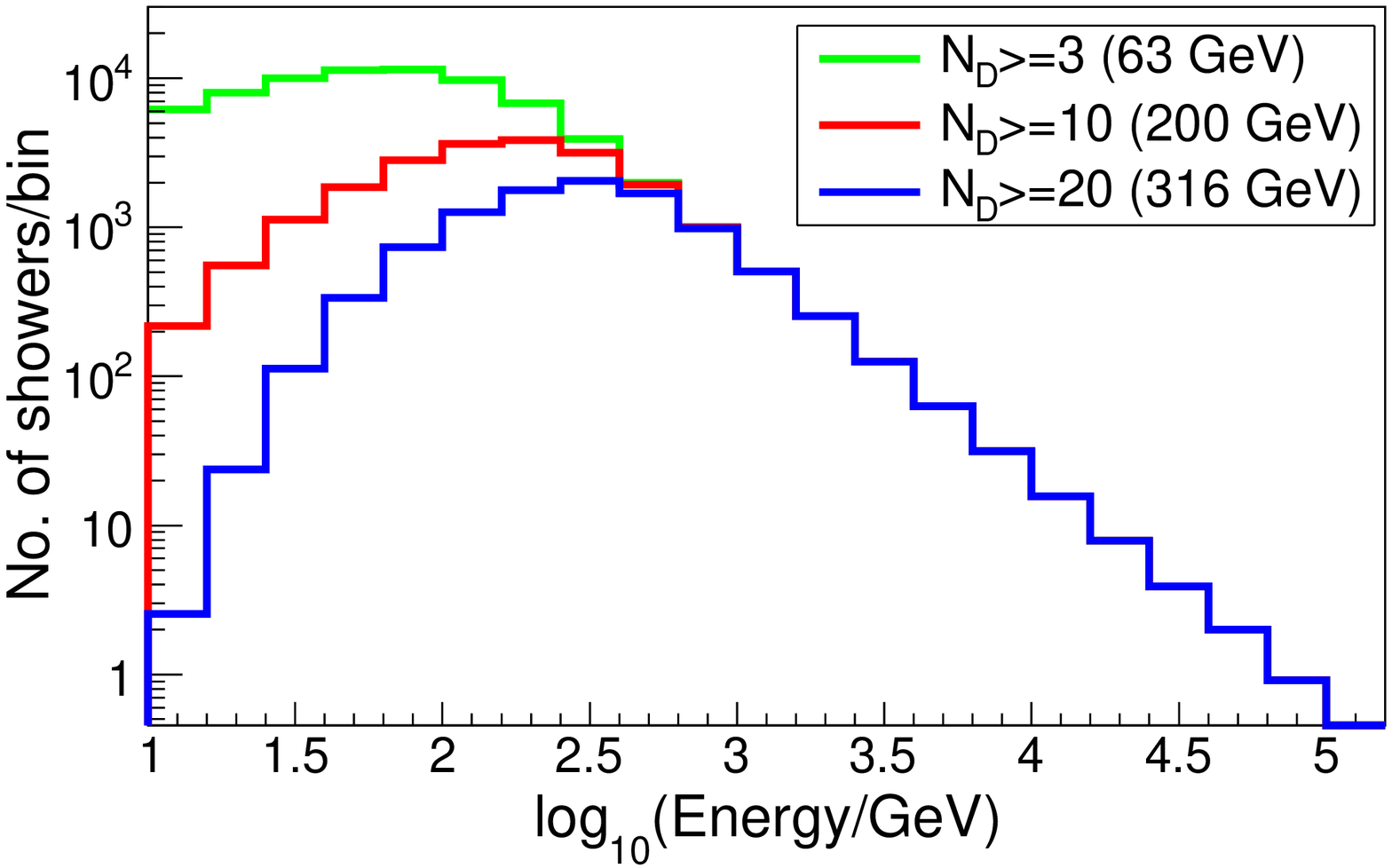}
\includegraphics[width=0.48\textwidth]{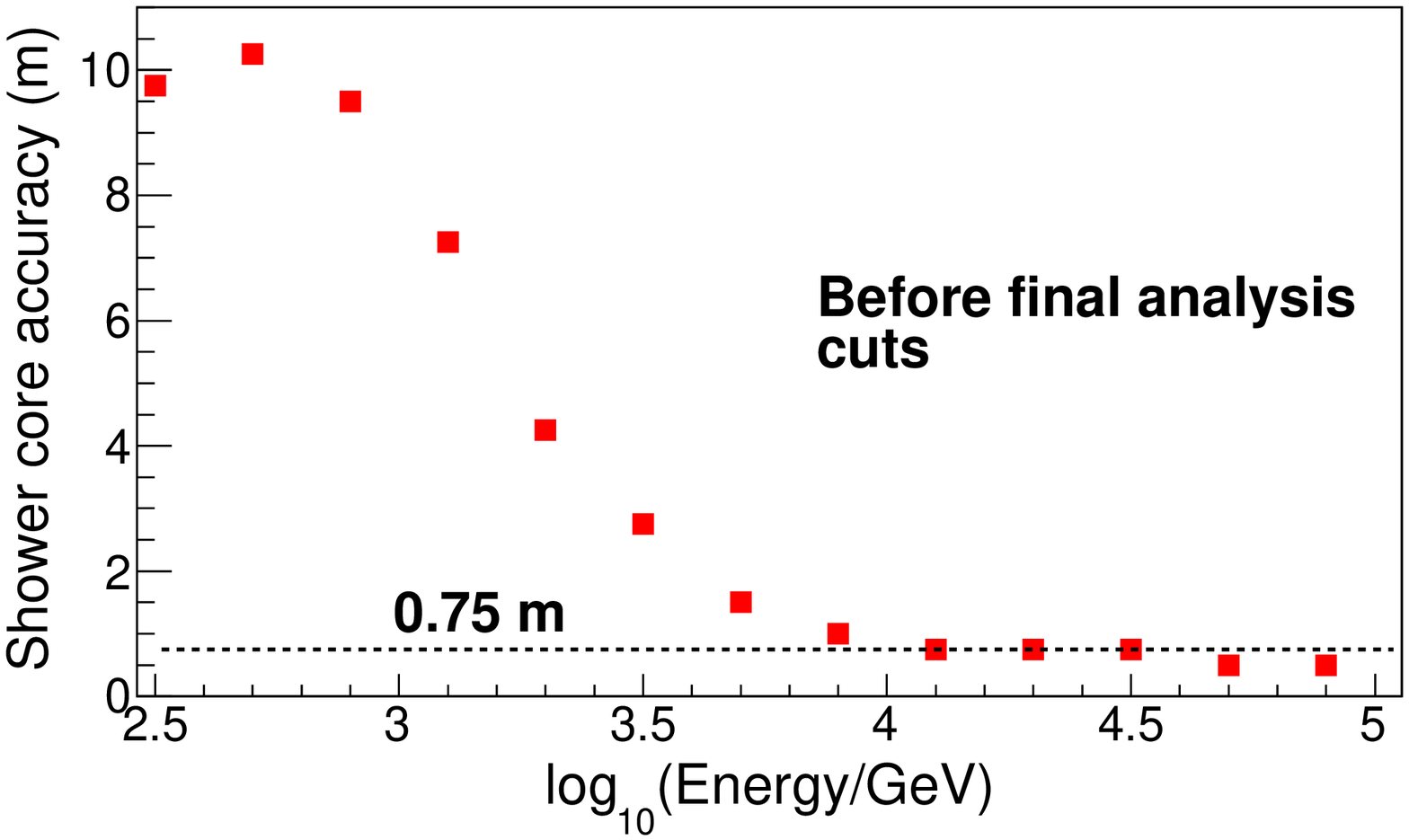}
\hspace{0.01cm}
\includegraphics[width=0.48\textwidth]{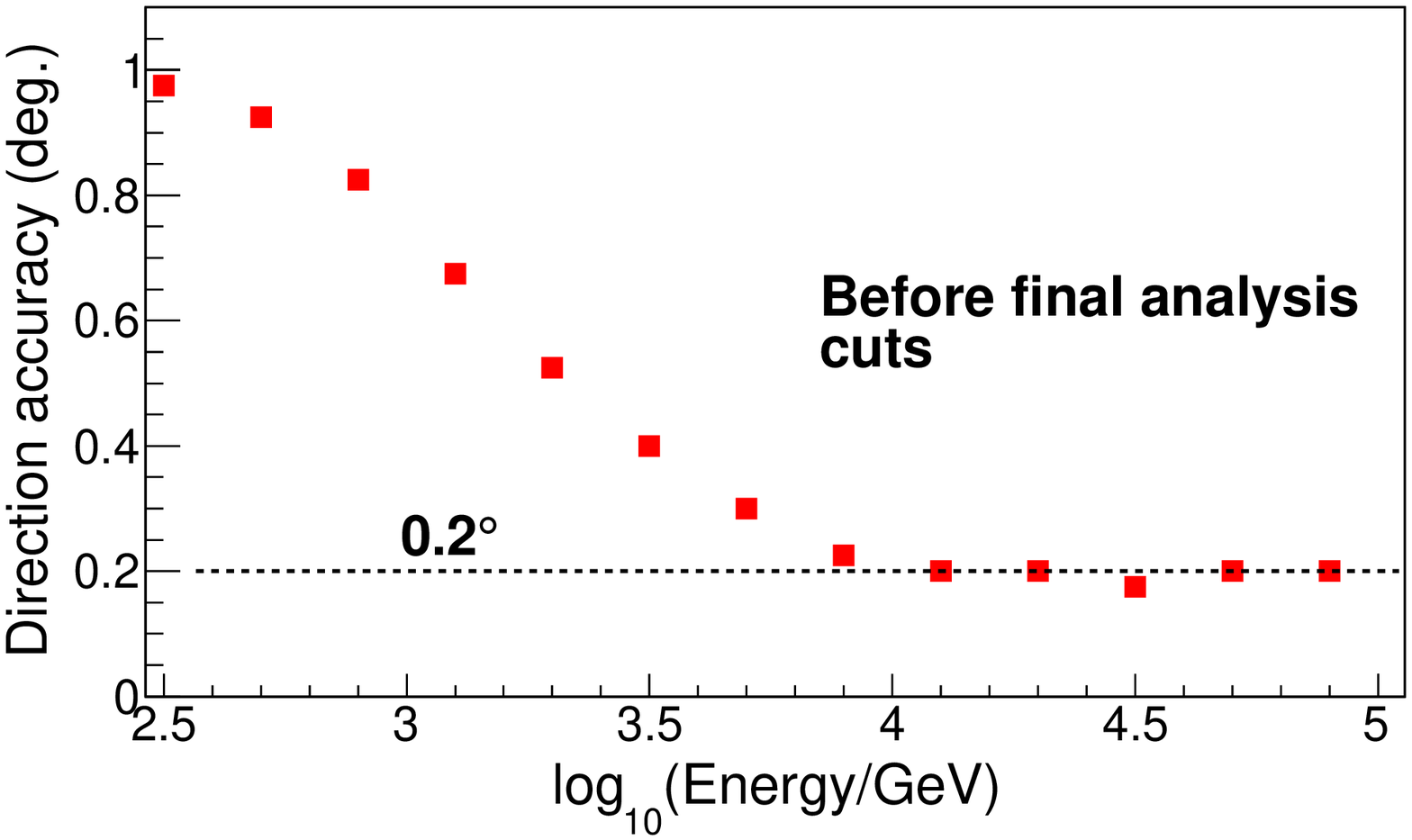}
\caption{\label{alto-param} \small{Energy distributions of gamma-ray showers (\textit{top-left}) after applying different trigger cuts based on the number of detectors: $\mathrm{N_D\geq(3, 10, 20)}$. The values given in the parentheses represent the peaks of the distributions. The other plots are reconstruction accuracies of the shower core (\textit{top-right}) and arrival direction (\textit{bottom}) for gamma-ray showers above the peak energy ($316$~GeV) for the case of $\mathrm{N_D\geq20}$. The results obtained are after a preliminary selection cut of 60~m in the reconstructed core, with no final analysis cut.}}
\end{figure}

For the performance study of the array, we simulate $\sim17$ million gamma-ray showers at a fixed zenith angle of $18^\circ$ in the energy range of 10~GeV$-$100~TeV assuming a differential energy spectrum of index $-2$. Each simulated shower is assigned a random core position within a radius of  $100$~m from the center of the array. Only showers with reconstructed core within 60~m are finally selected. Figure \ref{alto-param} (top-left) shows the energy distributions of the showers after applying different trigger cuts based on the number of WCDs, $\mathrm{N_D}$. Note that the showers are weighted so as to generate a "Crab-like" distribution of spectral index $-2.6$. The peaks of the distributions are also indicated in the figure. At a reasonable trigger condition of 10 detectors ($\mathrm{N_D}\geq10$), the peak value reaches $\sim\,200$~GeV.

Figure \ref{alto-param} also shows the reconstruction accuracies of the shower core (top-right) and arrival direction (bottom) as a function of gamma-ray energy for the case of $\mathrm{N_D}\geq20$. Above $\sim\,10$~TeV, the  reconstruction accuracy reaches $\sim\,0.75$~m for the shower core and $\sim\,0.2^\circ$ for the arrival direction. It should be noted that the results are obtained only after a preliminary selection cut of 60~m in the reconstructed core and no final analysis cut has been applied.

\section{Current and future activities}
The main design study of ALTO is at the final stage. Current efforts in simulation include study of the signal/background discrimination, and optimisation of the detector design and array configuration which include investigation of using slightly different sizes of PMTs as well as different concrete and scintillator thicknesses. The optimisation will be mainly driven by the goal to achieve an energy threshold below $200$~GeV, energy resolution below $30\%$, angular resolution of the order of $0.1^\circ$ at TeV range, and an improvement in the sensitivity by a factor of $5-10$ over existing similar experiments. Efforts are also underway to measure the optical properties of the liquid scintillator and the detector tanks for inclusion in further Monte-Carlo simulations. Preparations have also started for the construction of a prototype unit. Future work will involve testing with the prototype, and further fine tuning of the detector design based on results obtained from the prototype extrapolated using detailed simulations. Based on the final results, the electronics will need to be adapted to the required timing and waveform precisions, and coincidence trigger logic.

\section{Acknowledgements}
The ALTO project is being supported by the following Swedish private foundations or pub- lic institutes: the Crafoord Foundation, the Foundation in memory of Lars Hierta, the Magnus Bergvall's Foundation, the Crafoord stipendium of the Royal Swedish Academy of Sciences (KVA), the Märta and Erik Holmberg Endowment of the Royal Physiographic Society in Lund, the Läng- manska kulturfonden, the Helge Ax:son Johnson's Foundation and Linnaeus University. We also thank the Swedish National Infrastructure for Computing (SNIC) at Lunarc (Lund, Sweden).
We would like to thank Bertrand Vallage from CEA/Saclay (France) for providing us with two ANTARES optical modules. We would like to thank Jean-Pierre Ernenwein from CPPM/Marseille (France) for providing the ANTARES surface array and the scintillator liquid, for his extremely valuable expertise on scintillator materials characteristics, and for his novel ideas on the PMT installation and choice. Thanks also to Staffan Carius, Dean of the Faculty of Technology at Linnaeus University, for all the local support for the project.

\end{document}